\documentclass{aa}

\usepackage{graphicx}
\usepackage{wasysym}
\usepackage{amssymb}
\usepackage{amsmath}

\usepackage{graphicx} 
\usepackage[latin1]{inputenc} 

\begin{document}

\title{Accurate free and forced rotational motions of rigid Venus}

\author{L.~Cottereau \inst{1}, J.~Souchay\inst{2, 1}, S~Aljbaae\inst{3, 1}}


\institute{Observatoire de Paris, Syst\`emes de R\'ef\'erence Temps Espace (SYRTE), CNRS/UMR8630, Paris, France}

\date{}

\abstract
{The precise and accurate modelling of a terrestrial planet like Venus is an exciting and challenging topic, all the more interesting since it can be compared with that of the Earth for which such a modelling has already been achieved at the milliarcsecond level} 
{We want to complete a previous study (Cottereau and Souchay, 2009), by determining at the milliarcsecond level the polhody, i.e. the torque-free motion of the axis of angular momentum of a rigid Venus in a body-fixed frame, as well as the nutation of its third axis of figure in space, which is fundamental from an observational point of view.  }
{We use the same theoretical framework as Kinoshita (1977) did to determine the precession-nutation motion of a rigid Earth. It is based on a representation of the rotation of rigid Venus, with the help of Andoyer variables and a set of canonical equations  in Hamiltonian formalism}
{In a first part we have computed the polhody, i.e. the respective free rotational motion of the axis of angular momentum of Venus with respect to a body-fixed frame. We have shown that this motion is highly elliptical, with a very long period of 525 cy to be compared with 430 d for the Earth. This is due to the very small dynamical flattening of Venus in comparison with our planet. In a second part we have computed precisely the Oppolzer terms which allow to represent the motion in space of the third Venus figure axis with respect to Venus angular momentum axis, under the influence of the solar gravitational torque. We have determined the corresponding tables of coefficients of nutation of the third figure axis both in longitude and in obliquity due to the Sun, which are of the same order of amplitude as for the Earth. We have shown that the coefficients of nutation for the third figure axis are significantly different from those of the angular momentum axis on the contrary of the Earth. Our analytical results have been validated by a numerical integration which revealed the indirect planetary effects.}
{This paper is a complementary study of Cottereau and Souchay (2009).It  gives a precise determination both of the torque free motion of Venus, as well as the nutation of the third Venus figure axis in space for a short time scale, when considering the planet as a rigid body.}
\maketitle

\section{Introduction}
Venus which can be considered as the twin sister of the Earth, in view of its global characteristics (size, mass, density), has been the subject of a good amount of investigations on very long time scales, to understand its slow retrograde rotation (243 d) and its rather small obliquity ($2^{\circ}.63$) (Goldstein 1964; Carpenter 1964; Goldreich and Peale 1970; Lago and Cazenave 1979; Dobrovoskis 1980; Yoder 1995; Correia and Laskar 2001, 2003). Habibullin (1995) made an analytical study on the rotation of a rigid Venus. In Cottereau and Souchay (2009) we presented an alternative study, from a theoretical framework already used by Kinoshita (1977) for the rigid Earth. We made an accurate description of the motion of rotation of Venus at short time scale. We calculated the ecliptic coordinates of Venus orbital pole and the reference point $\gamma_{0V}$ which is the equivalent of the vernal equinox for Venus. Our value for the precession in longitude was $\dot{\Psi}=4474".35$t/cy$\pm 66.5$. We have performed a full calculation of the coefficients of nutation of Venus and presented the complete tables of nutation in longitude $\Delta \Psi$  and obliquity $\Delta \epsilon$, for the axis of angular momentum due to both the dynamical flattening and triaxiality of the planet.

In this paper, the study begun in Cottereau and Souchay (2009) is completed. First in section \ref{torque} we consider the torque free rotational motion of  a rigid Venus. We recall the parametrization of Kinoshita (1977) and the important equations of Kinoshita (1972) which are used to solve this torque free motion. The important characteristics (amplitude, period, trajectory) of the free motion are given . Cottereau and Souchay (2009) supposed that the relative angular distances between the three poles (of angular momentum, figure and rotation) are very small as it is the case for the Earth. In this paper we want to determine accurately the motion of the third Venus figure axis which is the fundamental one in an observational point of view. To do that we reject the hypothesis of coincidence of the poles. Thus we determine the Oppolzer terms depending on the dynamical flattening and the trixiality of Venus. Then we compare these terms with the corresponding terms of nutation for the axis of angular momentum, as determined by Cottereau and Souchay (2009) and the Oppolzer terms determined by Kinoshita (1977) for the Earth  in section \ref{section3}. We give the complete tables of the coefficients of nutation of the third figure axis of Venus. We compare them, with the coefficients of nutation of the angular momentum axis, taken in Cottereau and Souchay (2009) (section \ref{section4}). Finally in section \ref{section6} we determine the nutation of the angular momentum axis by numerical integration using the ephemeris DE405. We validate the analytical results of Cottereau and Souchay (2009) down  to a precision of the order of a relative $10^{-5}$. We show that the discrepancies between the numerical integration and analytical results (Cottereau and Souchay, 2009) are caused by the indirect planetary effect. i.e the small contribution to the nutation, which is due to the periodic oscillations of the orbital motion of Venus. In this paper as it was the case in Cottereau and Souchay (2009), our domain of validity is roughly 3000 years.

\section{Torque free motion for rigid Venus}\label{torque}

\subsection{Equation of torque-free motion}\label{equationTF}
\begin{figure}[htbp]
\center
\resizebox{1.\hsize}{!}{\includegraphics{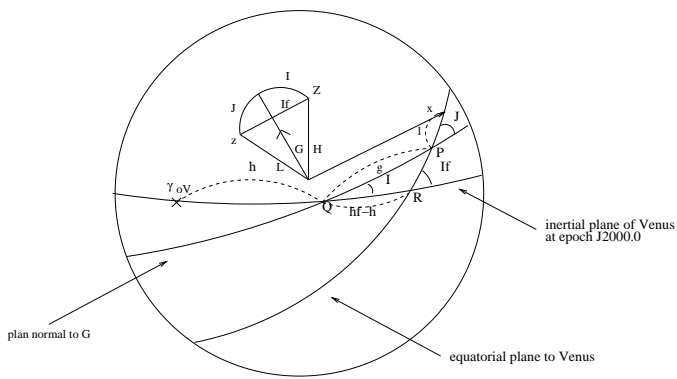}}
 \caption{Relation between the Euler angles and the
Andoyer variables.}
\label{fig1}
\end{figure}

We consider the problem of the rotational motion of the rigid Venus in absence of any external force. We note (0, X, Y, Z)  the inertial frame and (0,x,y,z) the cartesian coordinates fixed to the rigid body of the planet (see Fig.\ref{fig1}.). The orientation of Venus, with respect to the inertial axes, is determined through the Euler angles : $h_{f}$, $I_{f}$, $\phi$. The parameter $\phi$ gives the position of the prime meridian (0,x) with respect to $\gamma_{0V}$  (Kinoshita,1977). The angular momentum axis (hereafter denoted AMA) of Venus is the axis directed along $G$.
The components of the angular momentum referred to the body-fixed axes are :
\begin{eqnarray}
&&L_{x}=A\omega _{x}=A(\dot{h_{f}}\sin I_{f}\sin \phi+\dot{I_{f}}\cos\phi )\nonumber\\&&
L_{y}=B\omega _{y}=B(\dot{h_{f}}\sin I_{f}\cos \phi-\dot{I_{f}}\sin\phi )\nonumber\\&&
L_{z}=C\omega _{z}=C(\dot{h_{f}}\cos I_{f}+\dot{\phi})
\end{eqnarray}
where $A$, $B$, $C$ are the principal moments of inertia of Venus.
Moreover the kinetic energy is :
\begin{eqnarray}
T=\frac{1}{2}(\omega_{x}L_{x}+\omega_{y}L_{y}+\omega_{z}L_{z}).
\end{eqnarray}
To describe the torque free rotational motion we use the Andoyer variables  (Andoyer, 1923; Kinoshita, 1972)  (see Fig.\ref{fig1}.):
\begin{itemize}
\item $L$ the component of the angular momentum along the 0z axis
\item $H$ the component of the angular momentum  along the 0Z axis
\item $G$ the amplitude of the angular momentum of Venus
\item $l$ the angle between the origin meridian Ox and the node $P$
\item $h$ the longitude of the node of the AMA with respect to $\gamma_{0V}$
\item $g$ the longitude of the node of the plane (0, X, Y) with respect to Q and to the equatorial plane.
\end{itemize}
This parametrization is described in details in Kinoshita (1972, 1977) and in Cottereau and Souchay (2009). From these definitions we have :
\begin{eqnarray}\label{eq3}
L=G\cos J, \quad H=G\cos I
\end{eqnarray}
where $I$, $J$ are respectively the angle between the AMA and the inertial axis (O, Z), and the angle between the AMA and the third figure axis (hereafter denoted TFA). Using spherical trigonometry, we determine the following relation between the variables:
\begin{equation}
\phi=l+g
\end{equation}
The components of the angular momentum vector with the Andoyer variables are : 
\begin{eqnarray}
&&L_{x}=\sqrt{G^2-L^2}\sin l\nonumber\\&&
L_{y}=\sqrt{G^2-L^2}\cos l\nonumber\\&&
L_{z}=L.
\end{eqnarray}
The Hamiltonien for the torque-free motion of Venus corresponding to the kinetic energy is :
\begin{eqnarray}\label{hamiltonien}
H=\frac{1}{2}(\frac{\sin^{2} l}{A}+\frac{\cos^{2} l}{B})(G^2-L^2)+\frac{L^2}{2C}
\end{eqnarray}
The Hamiltonian $H$ does not depend on the time and it is free from $g$ and $h$. Thus the number of degrees of freedom of the torque free motion is one. Deprit (1967) has characterized this motion by studying the isoenergetic curves in the phase plane L-l. However we prefer to study the isoenergetic curves in the ($J$, $l$) phase space using the equation (\ref{eq3}).  This allows a good description of the position of the AMA with respect to the TFA. 
\begin{figure}[htbp]
\center
\resizebox{1\hsize}{!}{\includegraphics[angle=-90]{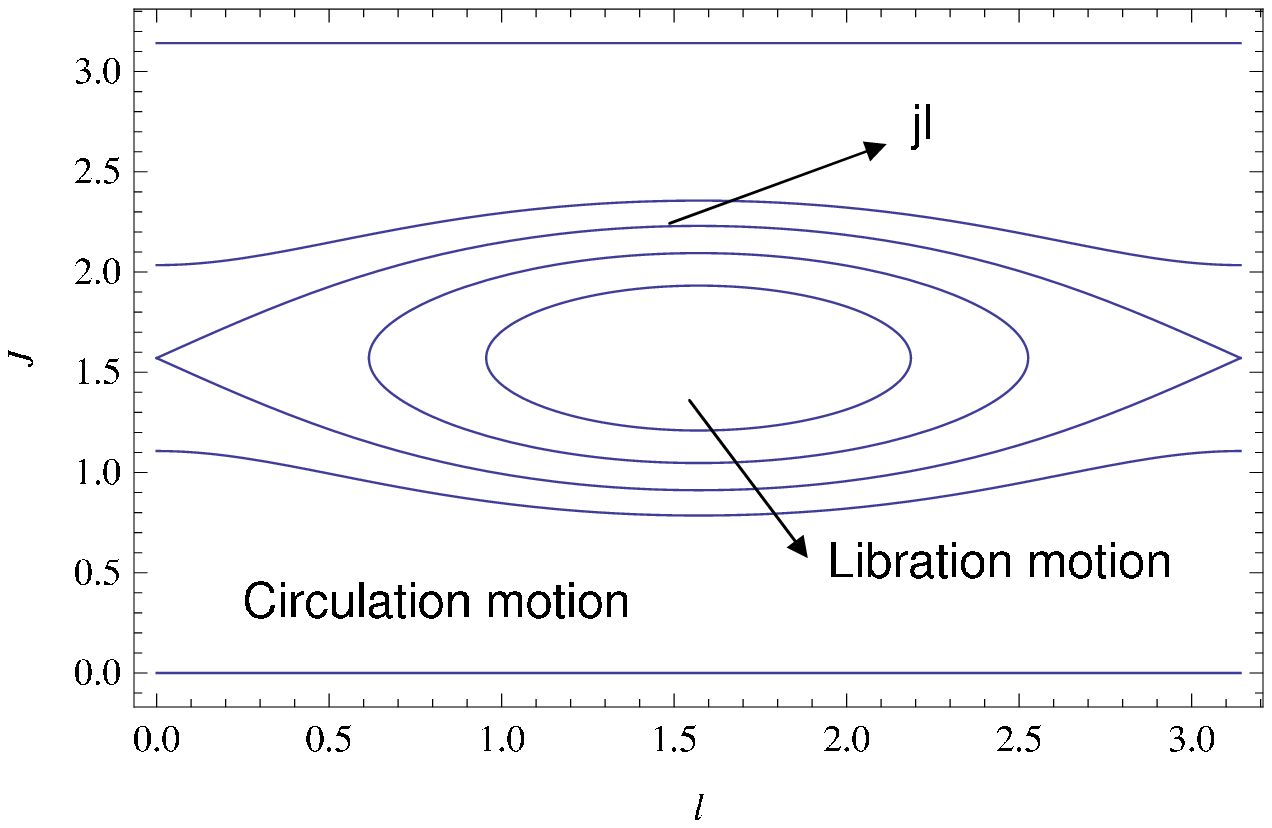}}
 \caption{Isoenergetic curves of Venus  in the ($J$, $l$) phase space of the torque free motion. Two motion are showed: "the libration motion" and the "circulation motion". }
\label{fig2}
\end{figure}
Fig.\ref{fig2} shows the two possible motions : one is the motion of libration and the other one the motion of circulation. Using the values of the relative differences of the moments of inertia of Williams (private communication):
\begin{eqnarray}\label{MOMI}
&&\frac{C-A}{MR^2}=5.519\times 10^{-6},\frac{C-B}{MR^{2}}=3.290\times 10^{-6},\nonumber\\&&\frac{B-A}{MR^{2}}=2.228\times 10^{-6}
\end{eqnarray}
and addopting $\frac{C}{MR^2}=0.3360$ (Yoder, 1995), we get  $\frac{C}{A}=1.000016$,  $\frac{C}{B}=1.000010$. We introduce here $j$ which corresponds  to the minimum value of J when $l=\frac{\pi}{2}$. We denote by $j_{l}$ the value reached  by $j$ on the separatrix. The libration motion is not possible if $J<j_{l}$ or $J>(\pi-j_{l})$.
The numerical values of $j_{l}$ can be determined using the equations  of Kinoshita (1972).  We have :
\begin{eqnarray}\label{jlimite}
j_{l}=\arccos{\sqrt{\frac{2e}{1+e}}}
\end{eqnarray}
with
\begin{eqnarray}\label{e}
e=\frac{1}{2}(\frac{1}{B}-\frac{1}{A})D
\end{eqnarray}
and
\begin{eqnarray}\label{1d}
 \frac{1}{D}=\frac{1}{C}-\frac{1}{2}(\frac{1}{B}+\frac{1}{A})
\end{eqnarray}
where $e$ measures the triaxiality of the rigid body (Andoyer, 1923).  Applying these formulas to Venus we get : $j_{l}=52^{\circ}.23$.

\subsection{Canonical transformations}\label{canT}

To determine the torque free motion of Venus, we use the method described in full details by Kinoshita (1972). We only give here the important results and equations which are needed to apply this study to Venus. The Hamiltonian for the torque free motion is given by (\ref{hamiltonien}). To solve the equations of motion, we perform a canonical transformation which replaces the angular momentum by its action variable (Goldstein,1964). First we make two intermediate transformations to simplify the computation.

\subsubsection{First transformation}

We change (L, G, l, g) to ($\alpha_{1}$, $\alpha_{2}$, $u_{1}$, $u_{2}$) using the Hamilton Jacobi method (Chazy, 1953). We note $S(L, G, \alpha_{1}, \alpha_{2}$) the characteristic function. We have the Hamilton-Jacobi equation :
\begin{equation}\label{rel3}
F_{0}= \frac{1}{2}(\frac{\sin ^{2} l}{A}+\frac{\cos ^{2} l}{B})((\frac{\partial S}{\partial g})^2-(\frac{\partial S}{\partial l})^2)+\frac{1}{2C}(\frac{\partial S}{\partial l})^2
\end{equation}
where
\begin{equation}
\alpha_{1}=F_{o},\ \alpha_{2}=\frac{\partial S}{\partial g}=G
\end{equation}
This equation can be solved as :
\begin{equation}\label{fonctionS}
S=\int \sqrt{\frac{\gamma^2-e\alpha_{2}^2 \cos 2l}{1-e\cos 2l}}dl+\alpha_{2}g
\end{equation}
where
\begin{equation}
\gamma^2=\big[2\alpha_{1}-\frac{1}{2}(\frac{1}{A}+\frac{1}{B})\alpha_{2}^2\big]D
\end{equation}
The constants $e$ and $D$ are given by (\ref{e}) and (\ref{1d}). Because the new Hamiltonian depends on only one of the momenta $\alpha_{1}$, we have : 
\begin{eqnarray}
&&u_{1}=t+\beta_{1}=\frac{\partial S}{\partial \alpha_{1}}=\frac{\partial S}{\partial \gamma}\frac{D}{\gamma}\nonumber \\&&
u_{2}=\beta_{2}=\frac{\partial S}{\partial \alpha_{2}}
\end{eqnarray} 

\subsubsection{Second transformation}

Then we make another transformation which changes ($\alpha_{1},  \alpha_{2}$, $ u_{1},  u_{2}$) to $(\bar{L}, \bar{G}, \bar{l}, \bar{g})$. The  two momenta are defined as follows :
\begin{eqnarray}\label{lbar}
&&\bar{L}=\gamma \nonumber \\&&
\bar{G}=\alpha_{2}=G
\end{eqnarray} 
With these new coordinates the Hamiltonian $F_{0}$ becomes :
\begin{equation}
F_{0}=\frac{1}{2D}\bar{L_{1}}^2+\frac{1}{4}(\frac{1}{A}+\frac{1}{B})\bar{G_{1}}^2
\end{equation}
Thanks to the Hamilton equations the conjugate variables are:
\begin{eqnarray}
&&\bar{l}=\frac{\partial F_{0}}{\partial \bar{L}}=\frac{\bar{L}}{D}(t+\beta_{1})\label{l1bar}\\&&
\bar{g}=\frac{\partial F_{0}}{\partial \bar{G}}=\frac{1}{2}(\frac{1}{A}+\frac{1}{B})\bar{G}(t+\beta{1})+\beta_{2}
\end{eqnarray}
Using (\ref{lbar}) we get $\bar{J}$ from:
\begin{eqnarray}
\bar{L}=\bar{G}\cos \bar{J}
\end{eqnarray}
and :
\begin{equation}\label{imp1}
\cos \bar{J}=\sqrt{1-(1+e) \sin^2 j}=\cos j \sqrt{1-e\tan^2 j}
\end{equation}
where $j$ is defined in the subsection \ref{equationTF}. Now we can introduce the canonical transformation which replaces the angular momentum by its action variable.

\subsubsection{Action variables}

The action variables are given by:
\begin{eqnarray}
&&\tilde{L}=\frac{1}{2\pi} \oint \sqrt{\frac{\bar{L}^2-e\bar{G}^2 \cos 2l}{1-e\cos 2l}}dl\nonumber\\&&
\tilde{G}=\bar{G}=G=\alpha_{2}
\end{eqnarray}
 $\tilde{l}$ and $\tilde{g}$ are canonically conjugate variables respectively to $\tilde{L}$ and $\tilde{G}$.
We make the following transformation to simplify the calculation: 
\begin{eqnarray}\label{rel4}
\cos 2\delta =\frac{\cos 2l-e}{1-e\cos 2l} \quad \mathrm{or} \quad \cos 2l =\frac{\cos 2\delta+e}{1+e\cos 2\delta}
\end{eqnarray}
We obtain :
\begin{eqnarray}\label{rel7}
&&\tilde{L}=\frac{\bar{G}}{2\pi}\frac{\sqrt{1-e^2}}{\sqrt{\bar{b}}}\oint \frac{\sqrt{1-\bar{k}^2\cos^2 \delta}}{1+e\cos 2\delta} d\delta \nonumber\\&&
=\bar{G} \wedge_{0},
\end{eqnarray}
where
\begin{eqnarray}
\bar{k}^2=\frac{2e}{1-e}(\bar{b}-1) \quad \bar{b}=(\cos^2 j)^{-1}
\end{eqnarray}
and
\begin{eqnarray}\label{gamma0}
\wedge_{0}=\frac{2}{\pi}\Big[E(k)F(\chi,k')+K(k)E(\chi,k')-K(k)F(\chi,k')\Big]
\end{eqnarray}
with
\begin{eqnarray}
\chi=\sin^{-1}\sqrt{\frac{1}{\bar{b}}}\quad k'=\sqrt{1-\bar{k}^2}.
\end{eqnarray}
In equation (\ref{gamma0}) $K(k)$ is the complete elliptic integral of the first kind with modulus $\bar{k}^2$, $E(k)$ that of the second kind. $F(\chi, k')$ is an incomplete elliptic integral of the first kind, $E(\chi, k')$ that of the second kind, $\wedge_{0}$ is a Heumann lambda function (Byrd and Friedman, 1954). The complete solution of  (\ref{rel7}) being of no interest here, we only give the mean motion of the angular variables $\tilde{l}$ and $\tilde{g}$ that will be used hereafter (Kinoshita, 1972, 1992). 
Thanks to the canonical transformations the time variation of $\tilde{l}$ and $\tilde{g}$ can be determined. We have :
\begin{eqnarray}
 \tilde{g}=\tilde{n_{g}}t+\beta_{2}=\tilde{n_{g}}t+\tilde{g_{0}}
\end{eqnarray}
\begin{eqnarray}
\tilde{l}=\tilde{n_{l}}t
\end{eqnarray}
where : 
\begin{eqnarray}\label{rel11}
\tilde{n_{l}}=\frac{\pi\bar{G}}{2KD}\sqrt{1-e^2}\cos j
\end{eqnarray}
\begin{eqnarray}\label{rel12}
\tilde{n_{g}}=\frac{G}{C}-\tilde{n_{l}}\wedge_{0}-G (\frac{1}{C}-\frac{1}{A}) \sin^2 j.
\end{eqnarray}
Here the epoch of time $t$ is defined such that $\tilde{l}=\frac{\pi}{2}$ at $t=0$.  Now we can give the development of the variables $g$ and $l$  which describe the torque free motion.
 
\subsection{Venus free rotation}\label{s3}
As a result we can use a development of our variables $g$, $l$ and $J$ with respect to $j$ and $e$. The results of these technical developments (Kinoshita, 1972) are summarized in the appendix.
As it is the case for the Earth we can suppose that the angle $j$ of Venus is very small. 
\begin{table}[!h]
\caption{ Important values of the free motion of Venus}
\begin{center}
\resizebox{0.5\hsize}{!}{\begin{tabular}[h]{lll}
\hline \hline
 Constant &numerical values  \\
\hline\\
$\frac{C}{A}$& 1.000016  \\

$\frac{C}{B}$ &1.000010  \\

$e$ & 0.230769  \\

$\frac{C}{D}$  & -0.000013  \\

$n_{l}$& 0.0119495 rd/cy\\

$T_{l}$&525.81 cy\\

$n_{g}$&0.0258549 rd/d\\

$T_{g}$&-243.02 d\\
\hline
\end{tabular}
}
\end{center}
\label{tab1}
\end{table}
Table \ref{tab1} gives the numerical values of the important constants used in the theory of the free motion of Venus. To calculate these values we take $\frac{C}{MR^2}$ (Yoder, 1995) and the values of the moment of inertia given in (\ref{MOMI}). The value of $D$ and $\tilde{b}$ are unknown. Finally we find:
\begin{eqnarray}
&&l=l^{*}-0.0576923 j^2 \frac{\sin 2\tilde{l}}{1+e\cos 2\tilde{l}} +O(j^4)\nonumber\\&&
g=\tilde{g}+\frac{G}{Dn_{\tilde{l}}}\big[-0.973009 (l^{*}-\tilde{l})\nonumber\\&&
+0.307692\times j^2[0.230769 \frac{\sin 2\tilde{l}}{1+e\cos 2\tilde{l}}\nonumber\\&&
+2.05548 (l^{*}-\tilde{l})]]+O(j^4)\nonumber\\&&
J=j\sqrt{1.28713+0.287129 \cos 2\tilde{l}}+O(e^3,j^2)\nonumber\\&&
\tan l^{*}=0.797007 \tan \tilde{l}.\nonumber
\end{eqnarray}  
Using the developments for a small $e$, we get:
\begin{eqnarray}
&&l=\tilde{l}-0.111538 \sin 2\tilde{l}+0.006221 \sin 4\tilde{l}+O(e^3, j^2)\nonumber\\&&
g=\tilde{g}+0.111538 \sin 2\tilde{l}-0.006221 \sin 4\tilde{l}+O(e^3, j^2)\nonumber\\&&
J=\tilde{J}[1.00933+0.11154 \cos 2\tilde{l}]+O(e^3, j^2)\nonumber\\&&
\tilde{J}=j+0.115385\tan j\nonumber\\&&
+0.05325\tan j (\frac{1}{8}+\frac{3}{16}\tan ^2 j)+O(e^3).
\end{eqnarray} 
The main limitation of our calculation being the uncertainty on the ratio $\frac{C}{MR^2}$ of Venus (Yoder, 1995), our polynomial expansions must be done accordingly. $4^{th}$ order terms are too small compared to our level of accuracy and have been discarded.
Projecting the pole of Venus on the (X, Y) plane with a value of $j=0.01$ rd for the developments above, we can plot the free motion of Venus (Fig.\ref{fig2}.).
\begin{figure}[htbp]
\center
\resizebox{0.7\hsize}{!}{\includegraphics{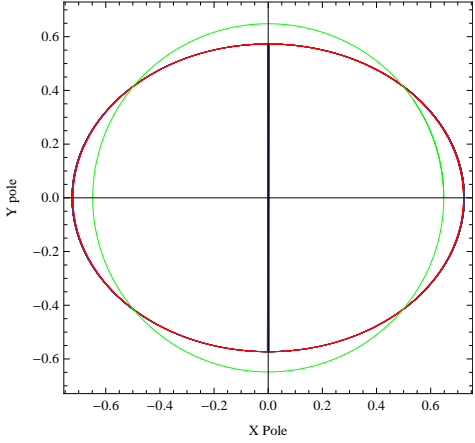}}
 \caption{(X, Y) free motion of Venus in space for five hundred century time space (red and blue curves). The green curve represents a circle with radius of  $r=\bar{j}$. We take $j=0.01$ rd.}
\label{fig3}
\end{figure}

We note here that the value of $j$ has been chosen arbitrarily, for it does not affect in a significant manner the polhody, except the amplitude. We see that the torque free motion of Venus is an elliptic motion, as it is the case for the Earth. The rotational free motion of Venus, with a period  $T_{l}=525$ centuries (Tab.\ref{tab1}) is much slower than the Earth one (303 d). If we consider the elastic Earth (ocean and atmosphere), the torque free motion has a period of 432 d, significantly larger than in the rigid case. The atmosphere of Venus is much denser than that of the Earth. So it would be interesting to study the torque free motion of the elastic Venus in a next paper. 

\section{Rigid Venus forced rotational motion}\label{section3}
In Cottereau and Souchay (2009) we supposed that the relative angular distances between the three poles of Venus (pole of angular momentum, of figure and of rotation) are very small, as it is the case for the Earth. In this section we want to determine the motion of the TFA, which is the fundamental one from an observational point of view. To do so we reject the hypothesis of coincidence of the poles. Using the spherical trigonometry in the triangle (P, Q, R) (see Fig.\ref{fig1}.) we determine the relations between the TFA and the AMA. Supposing that the angle $J$ between the AMA and the TFA is small we obtain:
\begin{eqnarray}
&&h_{f}=h+\frac{J}{\sin I} \sin g +O(J^2)\\&&\label{eqf1}
I_{f}=I+J \cos g+O(J^2)\label{eqf2}
\end{eqnarray}
where $I$ characterizes the obliquity and $h$ the motion of precession-nutation in longitude of the AMA of Venus.
$h_{f}$ and $I_{f}$ correspond to the same definitions as $h$ and $I$, but for the TFA  instead of the AMA. 
This yields (Kinoshita, 1977) :
\begin{eqnarray}
&&\Delta h_{f}=\Delta h+\Delta(\frac{J \sin g}{\sin I})+O(J^2)\label{eqf3}\\&&
\Delta I_{f}= \Delta I+ \Delta(J\cos g )+ O(J^2).\label{eqf4}
\end{eqnarray}
 where $\Delta h$ and $\Delta I$ represent the variation of the nutation of Venus, respectively in longitude and in obliquity.
The second terms in the right hand side of (\ref{eqf3}) and (\ref{eqf4}) are the so-called Oppolzer terms. They represent the difference between the nutation of the TFA and the nutation of the AMA (in longitude and in obliquity). These terms represent the differential effects of solid body tides on both axes.
Developing the equations (\ref{eqf3}) and (\ref{eqf4}) we obtain:
\begin{eqnarray}
&&\Delta(\frac{J \sin g}{\sin I})=\frac{1}{\sin I}(\Delta J \sin g +J\Delta g \cos g \nonumber\\&&
-\frac{J\sin g\Delta I\cos I}{\sin^2 I})\label{to1}\\&&
\Delta(J\cos g )=\Delta J\cos g-J\sin g\Delta g. \label{to2}
\end{eqnarray}
We see that (\ref{to1}) and (\ref{to2}) are  functions of $\Delta g$ and $\Delta J$. We remind here that $g$ and $J$ caracterize the motion of the AMA with respect to the TFA. As the Earth has a fast rotation the TFA and the AMA can be considered identical. Venus has a slow rotation, so it is interesting to see what difference will emerge on the motion of the TFA. To determine $\Delta g$ and $\Delta J$ we must solve the equations of motion.

\subsection{Equations of motion}
The Hamiltonian related to the rotational motion of Venus is (Cottereau and Souchay, 2009):
\begin{equation}
K"=F_{o}+ E + E'+U
\end{equation}
where $F_{o}$ is the Hamiltonian for the free motion, $E+E'$ is a component related to the motion of the orbit of Venus, which is caused by planetary perturbations. The expression of $F_{o}$ has been set in the previous section. The expression of  $E+E'$ is given in detail in Cottereau and Souchay (2009). $U$ is the disturbing potential due to the external disturbing body considered. Here the sole external disturbing body is the Sun (The perturbation due to the planets can be neglected in first order), and its disturbing potential is given by :
\begin{equation}\label{ddd}
U=\frac{\mathtt{\textbf{G}} M'}{r^3}[\frac{2C-A-B}{2}P_{2}(\sin \delta )+\frac{A-B}{4}P_2^{2} (\sin \delta) \cos 2\alpha]
\end{equation}
where $\mathtt{\textbf{G}}$ is the gravitationnal constant, $M'$ is the mass of the Sun, $r$ is the distance between  the Sun and Venus barycenters. $\alpha$ and $\delta$ are respectively the planetocentric longitude and latitude of the Sun, with respect to the mean equator of Venus and with respect to a meridian of origin (therefore $\alpha$ must not be confused with the usual right ascension). The $P_{n}^m$ are the classical Legendre functions given by :
\begin{equation}\label{legendre}
P_{n}^m (x)= \frac{(-1)^m(1-x^2)^{\frac{m}{2}}}{2^n n!}\frac{d^{n+m} (x^2-1)^n}{d^{n+m}x}.
\end{equation}
 In (\ref{ddd}) we only consider the potential in first order. The method for solving the equation of motion is described in Kinoshita (1977). Only the final results are given. We have :
\begin{eqnarray}
&&\Delta g=\frac{1}{G}(\cot J \frac{\partial W_{1}}{\partial J}+\cot I \frac{\partial W_{1}}{\partial I})\label{imp3}\\&&
\Delta J=\frac{1}{G}(\frac{1}{\sin J} \frac{\partial W_{1}}{\partial l}-\cot J \frac{\partial W_{1}}{\partial g})\label{imp4}
\end{eqnarray}
where
\begin{eqnarray}\label{form1}
&W_{1}&=\int \frac{\mathtt{\textbf{G}} M'}{r^3}\Bigg[\frac{2C-A-B}{2}P_{2}(\sin \delta )\nonumber \\&& +\frac{A-B}{4}P_{2}^{2} \sin \delta \cos 2\alpha \Bigg]dt.\nonumber \\
\end{eqnarray}
We use a transformation described by Kinoshita(1977) and based on the Jacobi polynomials. It expresses $\alpha$ and $\delta$ as  functions of $\lambda$ and $\beta$, respectively the longitude and the latitude of the Sun with respect to Venus mean orbital plane :
\begin{eqnarray}\label{aaa}
P_{2}(\sin \delta)&& = \frac{1}{2}(3\cos^2 J-1)\Bigg[\frac{1}{2}(3\cos^2 I-1)P_{2}(\sin \beta)\nonumber \\&&-\frac{1}{2} \sin 2I  P_{2}^1(\sin \beta)\cos 2(\lambda - h)\Bigg]\nonumber \\&&+\sin 2J \Bigg[-\frac{3}{4}\sin 2I P_{2}(\sin \beta)\cos g \nonumber \\ &&-\sum_{\epsilon =\pm 1}\frac{1}{4}(1+\epsilon \cos I)(-1+2\epsilon \cos I) \nonumber \\&& P_{2}^1(\sin \beta)\sin(\lambda-h-\epsilon g) \nonumber \\&& -\sum_{\epsilon =\pm 1}\frac{1}{8}\epsilon \sin I(1+\epsilon \cos I)\nonumber \\&& P_{2}^2(\sin \beta)\cos(2\lambda-2h-\epsilon g)\Bigg]+\sin^2 J \nonumber \\&& \Bigg[\frac{3}{4}\sin^2I P_{2}(\sin \beta)\cos 2g+\frac{1}{4}\sum_{\epsilon =\pm 1}\epsilon \sin I  \nonumber \\&& (1+\epsilon \cos I)P_{2}^1(\sin \beta)\sin(\lambda-h-2\epsilon g)-\frac{1}{16}\nonumber \\&& \sum_{\epsilon =\pm 1}(1+\epsilon \cos I)^2P_{2}^2(\sin \beta)\cos 2(\lambda-h-\epsilon g)\Bigg].\nonumber \\
\end{eqnarray}
and
\begin{eqnarray}\label{polynome}
&P_{2}^2(\sin \delta)&\cos 2\alpha =3\sin^2 J\Bigg[-\frac{1}{2}(3\cos^2 I-1)P_{2}(\sin \beta)\nonumber \\&& \cos 2l+\frac{1}{4}\sum_{\epsilon =\pm 1}\sin 2IP_{2}^1(\sin \beta)\sin(\lambda-h-2\epsilon l) \nonumber \\&& +\frac{1}{8}\sin^2 IP_{2}^2(\sin \beta)\cos 2(\lambda-h-\epsilon l)\Bigg]\nonumber \\ &&+\sum_{\rho =\pm 1}\rho \sin J (1+\rho \cos J)\nonumber \\&& \Bigg[-\frac{3}{2}\sin 2IP_{2}(\sin \beta) \cos (2\rho l+g)\nonumber \\&&-\sum_{\epsilon =\pm 1}\frac{1}{2}(1+\epsilon \cos I)(-1+2\epsilon \cos I)\nonumber \\&&  P_{2}^1(\sin \beta)\sin(\lambda-h-2\rho \epsilon l-\epsilon g)\nonumber \\&& -\sum_{\epsilon =\pm 1}\frac{1}{4}\epsilon \sin I(1+\epsilon \cos I) \nonumber \\&&  P_{2}^2(\sin \beta)\cos(2\lambda-2h-2\rho\epsilon l-\epsilon g)\Bigg].\nonumber
\end{eqnarray}
To simplify the calculations, we study separately the symmetric part of (\ref{form1}) depending on the dynamical flattening and the antisymmetric one depending on the triaxiality of Venus. So the coefficient of the dynamical flattening will be noted with a "s" index and the coefficient depending on the triaxiality with an "a" index. From its definition above, we can set: $\beta\approx 0$, for the latitude of the Sun with respect to Venus mean orbital plane can be considered as null.

\subsection{Oppolzer terms depending on the dynamical flattening}\label{oppolzerapl}

Using the equations (\ref{form1}) and (\ref{aaa}), with $\beta=0$, we have :
\begin{eqnarray}
W_{s1}&&= \frac{\mathtt{\textbf{G}} M'}{a^3}\frac{2C-A-B}{2}\int [(\frac{a}{r})^3P_{2}(\sin \delta)]dt\nonumber\\&&=
\frac{1}{2}(3\cos^2 J-1)W_{s10}-\frac{1}{2}\sin 2J W_{s11}\nonumber\\&&
+\frac{1}{4}\sin^2 J W_{s12} 
\end{eqnarray}
where
\begin{eqnarray}
&&W_{s10}=K_{s'}[-\frac{1}{6}(3\cos^2I-1) \int\frac{1}{2}(\frac{a}{r})^3dt\nonumber\\
&&-\frac{1}{4}\sin^2I\int\cos 2(\lambda-h)(\frac{a}{r})^3dt]
\end{eqnarray}
\begin{eqnarray}\label{ws11}
W_{s11}&&=K_{s'}[-\frac{1}{2}\sin 2I\int\frac{1}{2}(\frac{a}{r})^3\cos g \quad dt\nonumber\\&&
-\frac{1}{4}(1-\cos I)\int\cos (2\lambda-2h+g)(\frac{a}{r})^3 dt\nonumber\\&&
+\frac{1}{4}(1+\cos I)\int\cos (2\lambda-2h-g)(\frac{a}{r})^3 dt]
\end{eqnarray}

\begin{eqnarray}
W_{s12}&&=K_{s'}[-\sin ^2I\frac{1}{2}\int(\frac{a}{r})^3\cos 2g\quad dt\nonumber\\&&
-\frac{1}{4}(1-\cos I)^2\int\cos 2(\lambda-h+g)(\frac{a}{r})^3dt\nonumber\\&&
-\frac{1}{4}(1+\cos I)^2\int\cos 2(\lambda-h-g)(\frac{a}{r})^3dt]
\end{eqnarray}

and

\begin{eqnarray}
K_{s'}=\frac{3\mathtt{\textbf{G}}M'}{a^3}\frac{2C-A-B}{2}.
\end{eqnarray}
Thanks to the Hamilton equations, we obtain :
\begin{eqnarray}
\Delta_{s}g&&=\frac{1}{G}[-3\cos^2 J \ W_{s10}-\frac{\cos 2J\cos J}{\sin J}W_{s11}\nonumber\\&&
+\frac{1}{2}\cos^2 J \ W_{s12}]\nonumber\\&&
-\cos I \Delta h+O(J^2).
\end{eqnarray}
We suppose that the angle $J$ is small as is the case for the Earth.  This yields :
\begin{eqnarray}\label{deltag}
\Delta_{s} g&&=\frac{1}{G}[-3 W_{s0}-\frac{1}{ J}W_{s11}+\frac{1}{2} W_{s12}]\nonumber\\&&
-\cos I \ \Delta h+O(J^2).
\end{eqnarray}
We have also :
\begin{eqnarray}\label{deltaj}
&&\Delta_{s} J=\frac{1}{G}[\cos^2J \ \frac{\partial W_{s11}}{\partial g}-\frac{1}{8}\sin 2J \ \frac{\partial W_{s12}}{\partial g}]\nonumber\\&&
=\frac{1}{G}\frac{\partial W_{s11}}{\partial g}+O(J).
\end{eqnarray}
Using (\ref{deltag}) and (\ref{deltaj}) we obtain the Oppolzer terms depending on the dynamical flattening :
\begin{eqnarray}\label{op1}
&&\Delta_{s}(\frac{J \sin g}{\sin I})=\frac{1}{G \sin I}[\frac{\partial W_{s11}}{\partial g}\sin g\nonumber\\&&
-W_{s11} \cos g]+O(J)
\end{eqnarray}
\begin{eqnarray}\label{op2}
&&\Delta_{s}(J\cos g )=\Delta (J)\cos g-J\Delta g\sin g \nonumber\\&&
=\frac{1}{G}[\frac{\partial W_{s11}}{\partial g} cos g+W_{s11} \sin g]+O(J).
\end{eqnarray}

To solve the equations (\ref{op1}) and (\ref{op2}) through $W_{s11}$ given by (\ref{ws11}), it is necessary to develop $\frac{1}{2}(\frac{a}{r})^3\cos g$, $(\frac{a}{r})^3\cos (2\lambda-2h+g)$ and  $(\frac{a}{r})^3\cos (2\lambda-2h-g)$ with respect to the mean anomaly M,  the mean longitude of the Sun $L_{S}$ and g, the angle determined in Section \ref{torque}. Using the Kepler's law we obtain this development (See Table \ref{tdev1}, Table \ref{tdev2},Table \ref{tdev3}). Our value of the eccentricity was taken from Simon et al.(1994).

\begin{table}[!h]
\caption{Development of $ \frac{1}{2} \left(\frac{a}{r} \right) ^3  \cos(g)$ of Venus. $t$ is counted in Julian centuries.}
\label{tdev1}
\begin{center}
\resizebox{.8\hsize}{!}{\begin{tabular}[h]{lccrrr}

\hline \hline
    M & $L_{S}$ &g& period & &  \\
    & & &d& $\cos \times 10^{-7}$ & $t\cos \times 10^{-7}$   \\
\hline\\
 0 & 0 & 1 & -243.02    &$(\frac{1}{2}+\frac{3}{4}e^2)= 5000344$& -48    \\
\\
 1 & 0 & 1  & 2980.71  & $(\frac{3}{4}e+\frac{27}{32}e^3)=50792$  & -3582 \\
\\
 1 & 0 & -1 & 116.75  & $(\frac{3}{4}e+\frac{27}{32}e^3)=50792$  & -3582 \\
\\
 2 & 0 & 1& 208.948  & $(\frac{9}{8}e^2)=516$    & -72  \\
\\
 2 & 0 & -1& 76.83 & $(\frac{9}{8}e^2)=516$    & -72  \\
\\
3&0&1&108.27&$\frac{53}{16}e^3=10$&0\\
\\
3&0&-1&57.25&$\frac{53}{16}e^3=10$&0\\
\hline
\end{tabular}
}
\end{center}
\end{table}

\begin{table}[!hbtp]
\caption{Development of $ \left( \frac{a}{r} \right) ^3 \ \cos(2(\lambda-h) -g)$ of Venus. $t$ is counted in Julian centuries.}
\label{tdev2}
\begin{center}
\resizebox{.8\hsize}{!}{\begin{tabular}[h]{lcccrrr}
\hline \hline
  M & $L_{S}$ & g & period &&\\
 &  &  &d & $\cos \times 10^{-7}$ & $t \cos \times  10^{-7}$   \\
\hline \\
 0 & 2 & - 1 &76.83   & $(1-\frac{5}{2}e^2)=9998853$  & 161   \\
\\
 -1 & 2& -1  &116.75  & $(-\frac{1}{2}e+\frac{1}{16}e^3)=-33859$ & 2388 \\
\\
1 & 2 & - 1  & 57.25  &$(\frac{7}{2}e-\frac{123}{16}e^3)= 236993$ & -16718 \\
\\
 2 & 2 & - 1  & 45.62 & $(\frac{17}{2}e^2)=3898$ & -550\\
\\
-3&2&-1&-2980.71&$\frac{1}{48}e^3=0$&0\\
\\
3&2&-1&37.92&$\frac{845}{48}e^3=54$&-4\\
\hline
\end{tabular}
}
\end{center}
\end{table}
\newpage
\begin{table}[!hbtp]
\caption{Development of $ \left( \frac{a}{r} \right) ^3 \ \cos(2(\lambda-h) + g)$ of Venus. $t$ is counted in Julian centuries.}
\label{tdev3}
\begin{center}
\resizebox{.8\hsize}{!}{\begin{tabular}[h]{lcccrrr}
\hline \hline
 M & $L_{S}$ & g & period  &&\\
   &  &  &d &$\cos \times 10^{-7}$ & $t\cos \times 10^{-7} $ \\
\hline \\
 0 & 2 &  1  & 208.95 &$(1-\frac{5}{2}e_{V}^2)=9998853$ & 161   \\
\\
 -1 & 2& 1  & 2980.71  & $(-\frac{1}{2}e+\frac{1}{16}e^3)=-33859$ & 2388 \\
\\
 1 & 2 & 1 & 108.27 &$(\frac{7}{2}e-\frac{123}{16}e^3)= 236993$ & -16718 \\
\\
 2 & 2 &1 & 73.06  &  $(\frac{17}{2}e^2)=3898$ & 550\\
\\
-3&2&1&-116.75&$\frac{1}{48}e^3=0$&0\\
\\
3&2&1&55.14&$\frac{845}{48}e^3=54$&-4\\
\hline
\end{tabular}
}
\end{center}
\end{table}

The numerical value for $T_{g}$ is given in Table.\ref{tab1}. We remind here that our domain of validity is 3000 years as in the case in Cottereau and Souchay (2009). 
\begin{table}[!htbp]
\caption{Oppolzer terms in longitude depending on dynamical flattening [$\Delta \Psi_{s}=\Delta h_{s} $: nutation coefficients of the AMA].}
\label{ww1}
\begin{center}
\resizebox{1\hsize}{!}{\begin{tabular}[htbp]{crrrr}
\hline \hline
 Argument & Period & $\sin\omega t$ &$t\sin\omega t$ & $\cos\omega t$ \\
  & d & arc second &arc second/julian century &arc second \\
&& $(10^-7)$& $(10^-7)$& $(10^-7)$\\
\hline\\
$2L_{s}$ & 112.35&14962988[-21900468]  & -242[-352] & 0[0] \\

$M$ &224.70 &-6134235[889997] &432651[-62765] &5416[-61] \\

$ 2L_{s}+M$ & 74.90&264403 [-346057]&-18651 [24412] &-5[-7]\\

$2L_{s}-M$ & 224.70&-76066 [148323]&5366 [-10461] &2[-10]\\

$ 2L_{s}+2M$ & 56.17&3466 [-4269] & -489[602]&0[0] \\

$2M$ & 112.35& -5749[4521] & 811[-640] &0 [0] \\
\hline 

\end{tabular}
}
\end{center}
\end{table}

\begin{figure}[htbp]
\center
\resizebox{0.7\hsize}{!}{\includegraphics{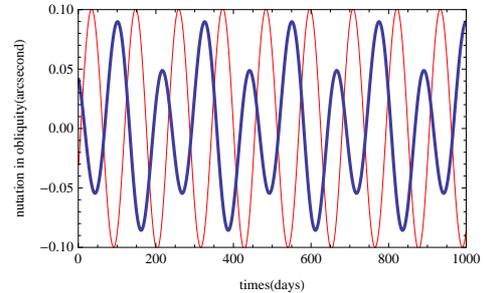}}
 \caption{The nutation of the AMA (red curve) and the nutation of the oppolzer terms (blue and bolt curve) in the obliquity of Venus depending on its dynamical flattening  for a 1000 d span, from J2000.0}
\label{fig4}
\end{figure}

\begin{figure}[htbp]
\center
\resizebox{0.7\hsize}{!}{\includegraphics{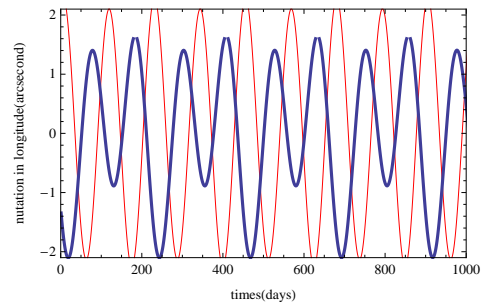}}
 \caption{The nutation of the AMA (red curve) and the nutation of the oppolzer terms (blue and bolt curve) in longitude of Venus depending on  its dynamical flattening  for a 1000 d span, from J2000.0}
\label{fig6}
\end{figure}

\begin{table}[!htbp]
\caption{Oppolzer terms in the obliquity depending on dynamical flattening [$\Delta \epsilon_{s}=\Delta I_{s}$ : nutation coefficients of the AMA].}
\label{ww2}
\begin{center}
\resizebox{1\hsize}{!}{\begin{tabular}[htbp]{crrrr}
\hline \hline
 Argument & Period & $\cos\omega t$ &$t\cos\omega t$ & $\sin\omega t$ \\
  &d & arc second &arc second/Julian century &arc second \\
&& $(10^-7)$& $(10^-7)$& $(10^-7)$\\
\hline\\
$2L_{s }$ & 112.35 &-690090 [100741]& -11[16] &0 [0] \\

$M$ &224.70 &-260831 [0] &1840[0] &-248 [0] \\

$2L_{s}+M$ & 74.90&-12184 [15919]& 859[-1123]&0[0] \\

$2L_{s}-M$ & 224.70& 3594[-6822]&-253 [481]& 0[0]\\

$2L_{s}+2M$ & 56.17&-160[196]&22[-27] &0[0] \\

$2M$ & 112.35& -122[0] &17 [0] &0 [0] \\
\hline
\end{tabular}
}
\end{center}
\end{table}
Thanks to the developments above we give in Table \ref{ww1} and Table \ref{ww2} the Oppolzer terms, respectively in longitude and in obliquity, depending on the dynamical flattening. For comparison we give the corresponding nutation coefficients (in brackets) of the AMA determined in Cottereau and Souchay (2009). They are represented in both Figs \ref{fig4} and \ref{fig6} for a 1000 d time span to see the leading oscillations. We can remark that the Oppolzer terms are of the same order of magnitude as the coefficients of nutation themselves. The Oppolzer terms associated with the argument $M$ are even larger.This is due to the small value of $\dot{M}+\dot{g}=-2\Pi/(-243.02/36525)+2\Pi/(224.70/36525)=76.99$ rd/cy which enters in the denominator during the integration of the equations of motion. Whereas for the calculation of the corresponding coefficient of the AMA only the numerical value $\dot{M}=2\Pi/(224.70/36525)= 1021.33$ rd/cy appears which is more larger than $\dot{M}+\dot{g}$.  In Figs \ref{fig4} and \ref{fig6} showing the oppolzer terms, the presence of the sinusoid with a period of 224 d reflect this fact .Remark (table \ref{ww2}) that the Oppolzer terms associated with the argument $M$ and $2M$ have a non zero amplitude, whereas the coefficients of nutation of the AMA associated with the same arguments do not exist. Indeed we remark here that to compute the coefficients of nutation in obliquity for this axis, we perform the derivative of $W_{1}$ with respect to $h$  which does not appear in the terms associated with the argument $M$ and $2M$. 
Notice also that for the Earth (see Kinoshita 1977), the Oppolzer terms depending on the dynamical flattening are negligible with respect to the coefficients of nutation of the AMA. The largest Oppolzer term in longitude in Kinoshita (1977) is $0".007559$ whereas for Venus  it is $1".4962$. In obliquity it is $0".002762$ whereas for Venus  it is $0".6901$. The rapid rotation of the Earth compared with the slow retrograde rotation of Venus explains this contrast : the frequencies which depend on the rotation $g$ and enter in the denominator during the integration are $10^4$ times larger for the Earth than for Venus.

Due to its slow rotation, the triaxiality of Venus ($1.66*10^{-6}$) is not negligible compared to the dynamical flattening ($1.31*10^{-5}$). Thus the Oppolzer terms depending on the triaxiality must be considered and are calculated in the following.

\subsection{Oppolzer terms depending on the triaxiality} 

Thanks to (\ref{form1}), we have :
\begin{eqnarray}
W_{a1}&&= \frac{\mathtt{\textbf{G}} M'}{a^3}\frac{A-B}{4}\int [(\frac{a}{r})^3P_{2}^2(\sin \delta) \cos 2\alpha]dt\nonumber\\&&
=\frac{3}{2}\sin^2 J W_{a10}-\sum_{\rho=\pm 1} \rho \sin J (1+\rho\cos J)W_{a11(\rho)}\nonumber\\&&
+\sum_{\rho=\pm 1}\frac{1}{4}(1+\cos J)^2 W_{a12(\rho)}
\end{eqnarray}
where
\begin{eqnarray}
W_{a10}&&=K_{a'}[(3\cos^2 I-1)\int \frac{1}{2}(\frac{a}{r})^3\cos 2l\quad dt\nonumber\\&&
+\frac{1}{4}\sin^2 I \int (\frac{a}{r})^3\cos 2(\lambda-h+l)dt\nonumber\\&&
+\frac{1}{4}\sin^2 I \int (\frac{a}{r})^3\cos 2(\lambda-h-l)dt\nonumber\\&&
\end{eqnarray}
\begin{eqnarray}
W_{a11(\rho)}&&=K_{a'}[-\frac{1}{2}\sin 2I\int (\frac{1}{2}\frac{a}{r})^3\cos (2\rho l+g)dt\nonumber\\&&
-\frac{1}{4}\sin I(1-\cos I)\nonumber\\&&
\int (\frac{a}{r})^3\cos (2\lambda-2h+2\rho l+g)dt\nonumber\\&&
+\frac{1}{4}\sin I(1+\cos I)\nonumber\\&&
\int (\frac{a}{r})^3\cos (2\lambda-2h-2\rho l-g)dt]
\end{eqnarray}
\begin{eqnarray}
W_{a12(\rho)}&&=K_{a'}[\sin^2I\int (\frac{1}{2}\frac{a}{r})^3\cos (2l+2\rho g)dt\nonumber\\&&
+\frac{1}{4}(1-\cos I)^2\int (\frac{a}{r})^3\cos 2(\lambda-h+\rho l+g)dt\nonumber\\&&
+\frac{1}{4}(1+\cos I)^2\nonumber\\&&
\int (\frac{a}{r})^3\cos 2(\lambda-h-\rho l-g)dt]
\end{eqnarray}
and :
\begin{eqnarray}
K_{a'}=\frac{3\mathtt{\textbf{G}}M'}{a^3}\frac{A-B}{4}.
\end{eqnarray}
Using the Hamilton equations and supposing, as for the terms depending on the dynamical flattening, that $J$ is a small angle, we obtain :
\begin{eqnarray}\label{deltaag}
\Delta_{Ag}&&=\frac{1}{G}[3W_{a10}-W_{a12(\rho)}-2 J W_{a11(1)}]\nonumber\\&&
-\cos I \ \Delta_{Ah}+O(J^2)
\end{eqnarray}
and :
\begin{eqnarray}\label{oaj}
\Delta_{aJ}&&=\frac{1}{G\sin J}\frac{\partial W_{a1}}{\partial l}-\frac{1}{G} \cot J\frac{\partial W_{a1}}{\partial g}\nonumber\\&&
=\frac{3}{2}\sin J \frac{\partial W_{a10}}{\partial l}\nonumber\\&&
-\frac{1+\cos J}{G}\Big[\frac{\partial W_{a11(1)}}{\partial l}-\cos J \ \frac{\partial W_{a11(1)}}{\partial g}\Big]\nonumber\\&&
+\frac{(1+\cos J)^2}{4G\sin J}\Big[\frac{\partial W_{a12(\rho)}}{\partial l}-\cos J \ \frac{\partial W_{a12(\rho)}}{\partial g}\Big].
\end{eqnarray}
As $W_{a11(-1)}$ is multiplied by $(1-\cos J)$, with our hypothesis it disappears from the equations (\ref{deltaag}) and  (\ref{oaj}). Since $W_{a11(1)}$ and $W_{a12(\rho)}$ include $l$ and $g$ in the form of  $g+2l$ and $2g+2l$ respectively, the last terms in the equation (\ref{oaj}) are negligible. Therefore $\Delta_{aJ}$ becomes:
\begin{eqnarray}
\Delta_{aJ}=-\frac{2}{G}\frac{\partial W_{a11(1)}}{\partial g}+ O(J).
\end{eqnarray}
The Oppolzer terms depending on the triaxiality are :
\begin{eqnarray}\label{eqt1}
&&\Delta(\frac{J \sin g}{\sin I})=-\frac{2}{G \sin I}[\frac{\partial W_{a11(1)}}{\partial g}\sin g\nonumber\\&&
+W_{a11(1)} \cos g]+O(J)
\end{eqnarray}
\begin{eqnarray}\label{eqt2}
&&\Delta(J\cos g )=\Delta J \cos g-J\Delta g\sin g \nonumber\\&&
=-\frac{2}{G}[\frac{\partial W_{a11(1)}}{\partial g} cos g-W_{a11(1)} \sin g].
\end{eqnarray}

To solve the equations (\ref{eqt1}) and (\ref{eqt2}) it is necessary to develop $\frac{1}{2}(\frac{a}{r})^3\cos (2l+g)$,   $(\frac{a}{r})^3\cos (2\lambda-2h+2l+g)$ and  $(\frac{a}{r})^3\cos (2\lambda-2h-2l-g)$ with respect to the mean anomaly M, the mean longitude of the Sun $L_{s}$ and the angles $l$ and $g$ determined in section (\ref{s3}). The coefficients are the same as those in Table \ref{tdev2} and \ref{tdev3}. Only the corresponding periods are different, because their calculation includes the argument $l$. The period $T_{l}$ is very long, as shown in Table \ref{tab1}.  Considering our level of accuracy, we suppose in this section that $e$ is constant and we take the value of Simon et al (1994) as $e=0.006771$.

We can determine the Oppolzer terms depending on the triaxiality.

\begin{table}[!htbp]
\caption{Oppolzer terms in longitude depending on triaxiality. Comparison with the corresponding nutation coefficients of the AMA in the tables of Cottereau and Souchay (2009)}
\label{ww3}
\begin{center}
\resizebox{.9\hsize}{!}{\begin{tabular}[htbp]{crrr}
\hline \hline
&&Oppolzer&A.M \\
 Argument & Period & $\sin(\omega t)$ &LC $sin(\omega t)$  \\
  & d & arc second & arc second  \\
&&$10^-7$&$10^-7$\\
\hline \\
$2\Phi$ & -121.51 &-11967515&5994459  \\

$2L_{S}-2\Phi$ &58.37&-3784751&2880826\\

$M+2\Phi$ & -264.6& 1490497&132590\\

$M-2\Phi+2L_{S}$&46.34& -66849&54201 \\

 $M-2\Phi$ & 78.86&58400 &-39519 \\

$2L_{S}+2\Phi$ &1490.35&5448 &-38866 \\

$-M-2\Phi+2L_{S}$&78.86& 19476&-13179\\

$2M+2\Phi$ & 1490.35&1062 &-7587  \\

$2M-2\Phi+2L_{S}$ &38.41&-876&739\\

$2M-2\Phi$&58.37&390&-297  \\

$M +2\Phi+2L_{S}$ & 195.26&67 &121 \\

$-M+2\Phi+2L_{S}$&-264.66&-263&-23 \\

$2M+2\Phi+2L_{S}$ &104.47&1&-1 \\

\end{tabular}

}
\end{center}
\end{table}

\begin{table}[!htpb]
\caption{Oppolzer terms in obliquity depending on triaxiality. Comparison with the corresponding nutation coefficients of the AMA in the tables of Cottereau and Souchay (2009).}
\label{ww5}
\begin{center}
\resizebox{0.9\hsize}{!}{\begin{tabular}[htbp]{crrr}
\hline \hline
&&Oppolzer&A.M \\
 Argument & Period & $\cos(\omega t)$&CS  $\cos(\omega t)$ \\
  & d & arc second&arc second\\
  &&$10^-7$&$10^-7$\\
\hline\\
$2\Phi$ & -121.51 &555905&-275453 \\

$2l_{S}-2\Phi$ &58.37&-173668&132365\\

$M +2\Phi$ & -264.6& -68544&-6093 \\

$M -2\Phi+2L_{S}$&46.34& 3074&2491\\

$2l_{S}+2\Phi$ &1490.35&-250&1786 \\

$M -2\Phi$ & 78.86&2686&1816\\

$-M -2\Phi+2L_{S}$&78.86& 896&-606\\

$2M +2\Phi$ & 1490.35&-49&348  \\

$2M -2\Phi+2L_{S}$& 38.41&-40&35\\

$2M-2\Phi$&58.37&18&14  \\

$M +2\Phi+2L_{S}$ & 195.26&-3&6 \\

$-M+2\Phi+2L_{S}$&-264.6&12&1 \\

$2M +2\Phi+2L_{S}$ &104.47&0&0 \\
\hline
\end{tabular}
}
\end{center}
\end{table}

\begin{figure}[htbp]
\center
\resizebox{0.7\hsize}{!}{\includegraphics{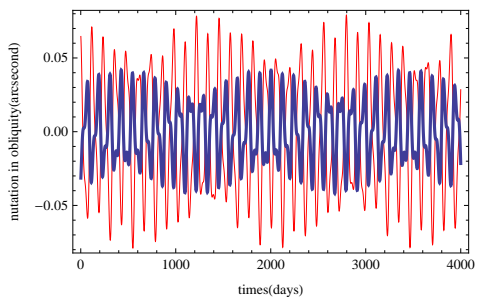}}
 \caption{The nutation of the AMA (Blue and bolt curve) and the nutation of the Oppolzer terms (red curve) in obliquity of Venus depending on its triaxility  for a 4000 d span, from J2000.0}
\label{fig5}
\end{figure}

\begin{figure}[htbp]
\center
\resizebox{0.7\hsize}{!}{\includegraphics{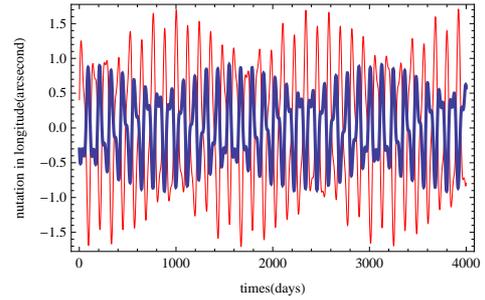}}
 \caption{The nutation of the AMA (Blue and bolt curve) and the nutation of the Oppolzer terms (red curve) in longitude of Venus depending on its triaxility  for a 4000 d span, from J2000.0}
\label{fig7}
\end{figure}

Table \ref{ww3} and Table \ref{ww5}  give the Oppolzer terms respectively in longitude and in obliquity, depending on Venus triaxiality. For comparison, we give the corresponding nutation coefficients of the AMA determined in Cottereau and Souchay (2009). They are represented in both Figs \ref{fig5} and \ref{fig7}, for a  4000 d time span. We remark here also that the Oppolzer terms are more important than the corresponding coefficients of nutation of the AMA The Oppolzer terms associated with the argument $2\Phi$ is even roughly twice larger than the corresponding coefficient of the nutation. This is due to the small value of  $\dot{2l}+\dot{g}=2\Pi/525.81-2\Pi/(243.02/36525)=944.36$ rd/cy which appears in the denominator during the integration of the equations of motion, whereas for the calculation of the corresponding coefficient of the AMA only the frequency of the sidereal angle $\dot{2\Phi}\approx \dot{2l}+\dot{2g}=1888.68$ rd/cy appears which is significantly larger. The appearance of the angle $\dot{g}$ during the integration explains that the other Oppolzer terms are more important than the corresponding coefficient of the nutation of the AMA, as justified in section \ref{oppolzerapl} for the terms depending on the dynamical flattening.

Oppositely to Earth, the Oppolzer terms in triaxiality are not negligible compared to the nominal values of the corresponding coefficient of nutation. The frequencies depending on the rotation $2l+g$ which enter in the denominator during the integration are very small compared those of our planet. Now we can determine the nutation of the TFA of Venus, which is fundamental from an observational point of view.

\section{Numerical results and comparison with the motion of the AMA}\label{section4}

Thanks to the equations (\ref{eqf3}) and (\ref{eqf4}) we calculate the coefficients of nutation of the TFA. We remind here that the nutation is respectively designated in longitude by $\Delta{h_{f}}$ and in obliquity by $\Delta{I_{f}}$. Tables \ref{co1} and \ref{co2} give the coefficients of nutation depending on the dynamical flattening respectively in longitude and in obliquity. In a similar way, Tables \ref{co3} and \ref{co4} give the coefficients depending on the triaxiality.

\begin{table}[!htbp]
\caption{$\Delta \Psi_{fs}=\Delta h_{fs} $ : nutation coefficients of the TFA in longitude of Venus depending on its dynamical flattening}
\label{co1}
\begin{center}
\resizebox{.9\hsize}{!}{\begin{tabular}[htbp]{crrrr}
\hline \hline
 Argument & Period & $\sin\omega t$ &$t\sin\omega t$ & $\cos\omega t$ \\
  & d & arc second &arc second/julian century &arc second \\
&& $(10^-7)$& $(10^-7)$& $(10^-7)$\\
\hline
$2L_{s}$ & 112.35& -6937480&-594 & 0 \\

$M$ &224.70 &-5244238&369886&5355\\

$ 2L_{s}+M$ & 74.90&-81654&5760&-2\\

$2L_{s}-M$ & 224.70&72257&-5095 &-8\\

$ 2L_{s}+2M$ & 56.17&-803 & 113&0[0] \\

$2M$ & 112.35& -1228 & 171&0 [0] \\
\hline

\end{tabular}
}
\end{center}
\end{table}

\begin{table}[!htbp]
\caption{$\Delta \epsilon_{fs}=\Delta I_{fs}$ : nutation coefficients of the TFA in obliquity of Venus depending on its dynamical flattening}
\label{co2}
\begin{center}
\resizebox{.9\hsize}{!}{\begin{tabular}[htbp]{crrrr}
\hline \hline
 Argument & Period & $\cos\omega t$ &$t\cos\omega t$ & $\sin\omega t$ \\
  &d& arc second &arc second/Julian century &arc second \\
&& $(10^-7)$& $(10^-7)$& $(10^-7)$\\
\hline\\
$2L_{s }$ & 112.35 &-589348& 5 &0 \\

$M$ &224.70 &-260831 &1839 &-248  \\

$2L_{s}+M$ & 74.90&3735&-264 & 0\\

$2L_{s}-M$ & 224.70&-3228 &-228& 0\\

$2L_{s}+2M$ & 56.17&37&-4 &0 \\

$2M$ & 112.35& -122&17 &0  \\
\hline
\end{tabular}
}
\end{center}
\end{table}

\begin{table}[!htbp]
\caption{$\Delta \Psi_{fa}=\Delta h_{fa} $  nutation coefficients of the TFA in longitude of Venus depending on its triaxialit}
\label{co3}
\begin{center}
\resizebox{.5\hsize}{!}{\begin{tabular}[htbp]{crr}
\hline \hline
 Argument & Period & $\sin(\omega t)$  \\
  & d & arc second  \\
&&$10^-7$\\
\hline \\
$2\Phi$ & -121.51 &-5973056  \\

$2L_{S}-2\Phi$ &58.37&-903925 \\

$M+2\Phi$ & -264.6& 1623087\\

$M-2\Phi+2L_{S}$&46.34&-12648 \\

 $M-2\Phi$ & 78.86&29199 \\

$2L_{S}+2\Phi$ &1490.35&-33418\\

$-M-2\Phi+2L_{S}$&78.86& 6297\\

$2M-2\Phi+2L_{S}$ &38.41&-137]\\

$2M-2\Phi$&58.37&195  \\

$M +2\Phi+2L_{S}$ & 195.26&188\\

$-M+2\Phi+2L_{S}$&-264.66&-286 \\

$2M+2\Phi+2L_{S}$ &104.47&0\\
\hline
\end{tabular}

}
\end{center}
\end{table}

\begin{table}[!htpb]
\caption{$\Delta \epsilon_{fa}=\Delta I_{fa} $ : nutation coefficients of the TFA in obliquity of Venus depending on its triaxiality }
\label{co4}
\begin{center}
\resizebox{0.5\hsize}{!}{\begin{tabular}[htbp]{crr}
\hline\hline
 Argument & Period & $\cos(\omega t)$  \\
  & d & arc second\\
\hline\\
$2\Phi$ & -121.51 &280452 \\

$2l_{S}-2\Phi$ &58.37&-41303 \\

$M +2\Phi$ & -264.6& -74637 \\

$M -2\Phi+2L_{S}$&46.34& 5565\\

$2l_{S}+2\Phi$ &1490.35&1536\\

$M -2\Phi$ & 78.86&4502\\

$-M -2\Phi+2L_{S}$&78.86& 290\\

$2M +2\Phi$ & 1490.35&299\\

$2M -2\Phi+2L_{S}$& 38.41&-5 \\

$2M-2\Phi$&58.37&32  \\

$M +2\Phi+2L_{S}$ & 195.26&3 \\

$-M+2\Phi+2L_{S}$&-264.6&13\\

$2M +2\Phi+2L_{S}$ &104.47&0 \\
\hline
\end{tabular}
}
\end{center}
\end{table}

\begin{figure}[htbp]
\center
\resizebox{0.7\hsize}{!}{\includegraphics{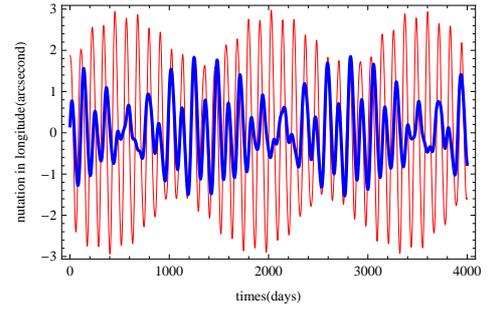}}
 \caption{The nutation of the TFA (blue curve) and the nutation of the momentum axis in longitude of Venus for 4000 d time span, from J2000.0}
\label{fig8}
\end{figure}

\begin{figure}[htbp]
\center
\resizebox{0.7\hsize}{!}{\includegraphics{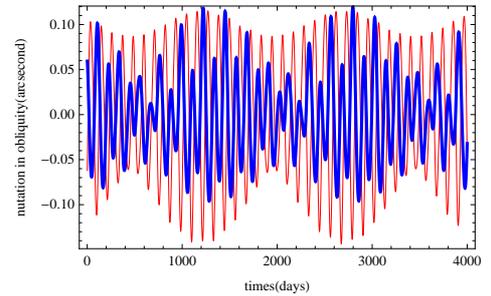}}
 \caption{The nutation of the TFA (blue curve) and the nutation of the momentum axis in obliquity of Venus for 4000 d time span, from J2000.0}
\label{fig9}
\end{figure}

In this section, we will show the difference between the nutation of the TFA, calculated in this paper and that of the AMA of Venus ($\delta{I}$ and $\Delta{h}$), as calculated by Cottereau and Souchay (2009). Figs \ref{fig8} and  \ref{fig9} represent respectively the nutation in longitude and in obliquity of the two axes for a 4000 d time span.

In longitude we can point out two important specific remarks :
\begin{itemize}
\item 
The nutation of the TFA is significantly smaller than the nutation of the AMA. Indeed the amplitude peak to peak, of the nutation of the TFA (an amplitude of 1.5") is twice smaller than that of the AMA (an amplitude of 3")
\item
The nutation of the TFA is dominated by three sinusoids associated with the arguments $2L_{s}$, $M$ and $2\Phi$, with respective periods 112.35 d, 224.70 d and 121.51 d whereas the nutation of the AMA is dominated by two sinusoids of argument  $2L_{s}$ and $2\Phi$. 

\end{itemize}
The same kind of remark is available in obliquity but the difference between the nutations is relatively less important. The amplitude of the nutation of the TFA is varying between 0".10 and 0".08 (peak to peak) whereas the amplitude of the AMA is varying between -0."14 and 0."11. The nutation in obliquity is also dominated by a sinusiod, with a period $M$ which is not as large as the corresponding sinusoid of the AMA.

Finally we can highlight the differences between the Earth and Venus. For the Earth, the nutation of the two axes (angular momentum and third figure axis) are roughly the same (Woolard,1953, Kinoshita, 1977) whereas in the case of Venus they are significantly different.  We can also remark that the leading component of nutation of the third Venus figure axis in longitude due to the gravitational action of the Sun, with argument $2L_{s}$ (see \ref{co1}) has an amplitude of  0".693. This is of the same order as the leading $2 L_{s}$  the nutation amplitude of the Earth  due to the Sun, i.e. 0".998 despite the fact that Venus has a very small dynamical flattening. As explained by Cottereau and Souchay (2009) this is due to the compensating role of the very slow rotation of Venus. Moreover, notice that in the case of Venus the argument $L_{s}$ stands for the longitude of the Sun as seen from the planet, so that the corresponding period of the leading nutation term with  $2L_{s}$ argument is 112.35 d, whereas it is 182.5 d in the case of the Earth. 

\section{Determination of the indirect planetary effects on the nutation of Venus}\label{section6}

\begin{figure}[htbp]
\center
\resizebox{0.7\hsize}{!}{\includegraphics{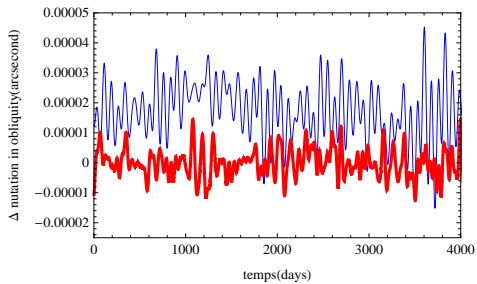}}
 \caption{Difference between the nutation in obliquity of the AMA and the numerical integration for 4000 d time span, from J2000.0. The curve at bottom represents the residual after substracting the sinusoidal terms of the Table \ref{planeff1}}
\label{fig12}
\end{figure}

\begin{figure}[htbp]
\center
\resizebox{0.7\hsize}{!}{\includegraphics{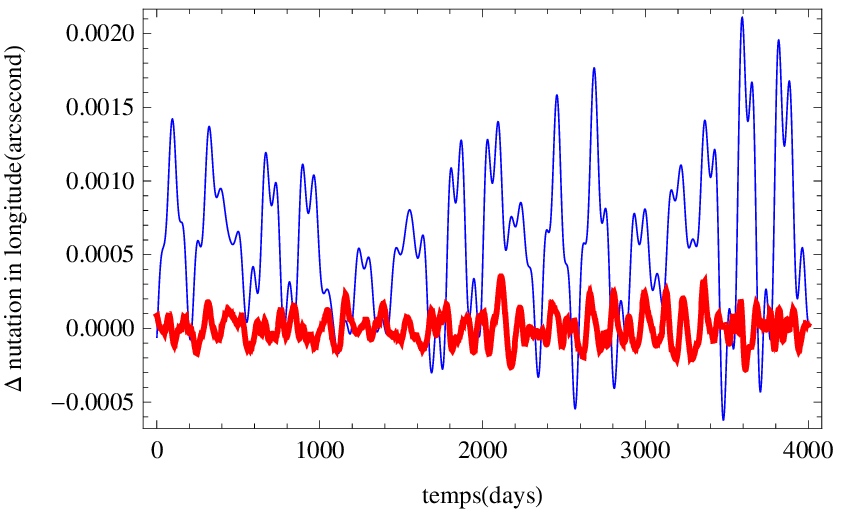}}
 \caption{Difference between the nutation in longitude of the AMA and the numerical integration for 4000 d time span, from J2000.0. The curve at bottom represents the residual after substracting the sinusoidal terms of the Table \ref{planeff2}.}
\label{fig13}
\end{figure}

Using the ephemeris DE405 (Standish, 1998), we have computed the nutation of Venus by numerical integration with a Runge-Kutta $12^{th}$ order algorithm. Figs \ref{fig12} and \ref{fig13} show the differences between the nutations in obliquity and in longitude of the AMA as computed from the analytical tables (Cottereau and Souchay, 2009) and that from the numerical integration for a 4000 d time span. The residuals obtained consist clearly in periodic components with small amplitudes, of the order of $10^{-5}"$ in obliquity and $10^{-3}"$ in longitude. This numerical integration validates the results of Cottereau and Souchay (2009), down to a relative accuracy of $10^{-5}$. Moreover Kinoshita's model used in Cottereau and Souchay (2009) supposed a Keplerian motion of Venus around the Sun. It is well known that the effects of planetary attraction into Earth's orbit (called indirect planetary effect) entails a departure from the Keplerian motion and that this departure induces new nutation terms as calculated by Souchay and Kinoshita (1996). In order to infer whether the discrepancies between our numerical integration and our analytical computation are caused by this indirect planetary effect in Venus orbit, we performed a spectral analysis of the residuals.

\begin{table}[!htpb]
\caption{ Coefficient of rigid Venus nutation from the indirect planetary contribution in longitude. Comparison with the respective value in tables of Souchay and Kinoshita (1995) when we have the Earth as a indirect planetary contribution.}
\label{planeff1}
\begin{center}
\resizebox{0.9\hsize}{!}{\begin{tabular}[htbp]{crrrrr}

\hline \hline
  Planetary effects & Period & amplitude&Amplitude&SK& SK   \\
  &&sin&cos&sin&cos\\
  & yr &0.01mas&0.01mas& 0.01mas&0.01mas\\
\hline\\
 2V-2E& 0.80 &31.5&-9.6&-9.6&0.0 \\

2V-3M&0.60&19.3&4.0&/&/ \\

3V-3E& 0.53& -2.0&13.4&0.5&-0.2 \\

V-J&0.64& 15.1&-26.2&/&/\\

 V-E&1.60&-2.6&-15.3&6.6&0.0\\

 & 4.11&-13.3&-38.4&/&/\\

2V+2M+2J&0.22& 6.2&12.7&/&/\\

& 9.31&-274.5&421.0&/&/\\

& 0.20&-1.5&-2.3&/&/ \\

&0.70&-21.0&-6.5&/&/  \\

2V-2Me& 0.19&8.5&-7.1&/&/ \\

2V+J+M&0.25&4.6&-2.9&/&/\\
\hline

\end{tabular}
}
\end{center}
\end{table}

\begin{table}[!htpb]
\caption{ Coefficient of rigid Venus nutation from the indirect planetary contribution in obliquity. Comparison with the respective value in tables of Souchay and Kinoshita (1995) when we have the Earth as a indirect planetary contribution.}
\label{planeff2}
\begin{center}
\resizebox{0.9\hsize}{!}{\begin{tabular}[htbp]{crrrrr}

\hline \hline
  Planetary effects & Period & amplitude&amplitude&SK& SK   \\
  &&sin&cos&sin&cos\\
  & yr &0.01mas&0.01mas& 0.01mas&0.01mas\\
\hline\\
 2V+2M+2J& 0.22&0.5&-0.4&/&/ \\

2S&14.96&0.7&0.4&/&/ \\

&4.06& -0.3&0.2&/&/\\

2V-2Mer&0.19&-0.3&-0.3&/&/\\

V-2J+M & 0.50&-0.3&0.0&/&/\\
\hline
\end{tabular}
}
\end{center}
\end{table}
The leading oscillations of the two signals (in longitude and in obliquity) are determined thanks to a fast Fourier Transform (FFT). Tables \ref{planeff1} and  \ref{planeff2} give the leading amplitudes and periods of the sinusoids characterizing the signal in Figs \ref{fig12} and \ref{fig13} where the curve at the bottom represents the residuals after subtraction of these sinusoids. The periods presented in the Tables \ref{planeff1} and  \ref{planeff2} do not correspond to any period of the tables given in the precedent section starting from the keplerian approximation. On the opposite, when comparing them with the tables of nutation of the Earth taken from Souchay and Kinoshita (1996), similar periods appear which correspond to the combination of planetary longitudes. This is a clear confirmation that the differences between our analytical computation and our numerical integration, in Figs \ref{fig12} and \ref{fig13}, are essentially due to the indirect planetary effects, negligible at first order and not taken previously into account by Cottereau and Souchay (2009). Tables \ref{planeff1} and \ref{planeff2} also present, when they have been clearly identified, the combination of the planetary longitudes corresponding to the detected sinusoids. Moreover, we give, when they are available, the corresponding amplitudes of the nutation of the Earth due to the indirect effect of Venus. We can thus point out the similitude and reciprocity of the indirect planetary effects of Venus on the rotation of the Earth, and of the indirect planetary effects of the Earth on the rotation of Venus.

\section{Conclusion and Prospects}

In this paper we achieved the accurate study of the rotation of Venus, for a rigid model and on a short time scale begun by Cottereau and Souchay (2009), by applying analytical formalisms already used for the rigid Earth (Kinoshita, 1972, 1977). The differences between the rotational characteristics of Venus and our planet, due to the slow rotation of Venus and its small obliquity, have been highlighted. 

Firstly we have precisely determined the polhody, i.e the torque free rotational motion for a rigid Venus. We have adopted the theory used by Kinoshita (1972, 1992) and we have given the parametrization and the equations of motion to solve the motion. We have shown that the polhody is significantly elliptic, on the contrary of the Earth for which it can be considered as circular in first approximation. Moreover it is considerably slower. Indeed, the period of the torque free motion is $525.81$ cy for Venus whereas it is $303$ d for our planet, when considered as rigid. 

Then we have determined the motion of the third figure axis, which is fundamental from an observational point of view. We have calculated the Oppolzer terms due to the gravitational action of the Sun using the equation of motion of Kinoshita (1977) as well as the corresponding development of the disturbing functions. We have compared them with the coefficients of nutation for the angular momentum axis taken in Cottereau and Souchay (2009). One of the important results is that these Oppolzer terms depending on the dynamical flattening are of the same order of amplitude as the coefficients of nutation themselves, whereas for the Earth (Woolard, 1953; Kinoshita, 1977) these Oppolzer terms are very small with respect to the coefficients of nutation of the angular momentum axis. Moreover we have computed the Oppolzer terms depending on the triaxiality, which is not done in Kinoshita (1977) in the case of the Earth for which they are negligible. In the case of Venus,these Oppolzer terms are significant. Even larger than the corresponding coefficients of nutation of the angular momentum axis. 

Thanks to our Oppolzer terms we have also been able to give the tables of nutation of the third figure axis from which we computed the nutation for a 4000 d time span. The comparison with the nutation of the angular momentum axis, calculated from Cottereau and Souchay (2009), is also given. The nutation of the third figure axis is significantly smaller peak to peak than the nutation of the angular momentum axis in longitude, and less important in obliquity. The amplitude of the largest nutation coefficient in longitude of the third figure axis (1.5")  is half the one of the angular momentum axis (3"). The amplitude of the nutation in obliquity of the third  figure axis is 0".18 peak to peak whereas the amplitude of the angular momentum axis is 0".25. The nutations of the third figure axis, in obliquity and in longitude, are dominated by three sinusoids associated with the arguments $2L_{s}$, $M$ and $2\Phi$, with respective periods 112.35 d, 224.70 d and 121.51 d. The nutation of the angular momentum axis is dominated by two sinusoids with the argument  $2L_{s}$ and $2\Phi$.  Our results have shown that although the axis of angular momentum and the third figure axis can be considered identical in the case of the Earth (Kinoshita, 1977), this approximation does not hold in the case of a slowly rotating planet as Venus.

At last we have validated our analytical results down to a relative accuracy of $10^{-5}$ with a numerical integration. Moreover, we have confirmed, by using results in Souchay and Kinoshita (1996) for the nutation of a rigid Earth, that the differences between our analytical computation and our numerical integration are essentially due to the indirect planetary effects, which was not taken into account by Cottereau and Souchay (2009). We think that this study is fundamental to understand the behavior of Venus rotation in a very accurate and exhaustive way for short time scales, and should be a necessary starting point to another similar study including non rigid effects (elasticity, atmospheric forcing etc...).

\section{appendix}

\subsection{Development for a small value of the triaxiality (Kinoshita, 1972)}

We note $\tilde{b}=\sqrt{\frac{G}{\tilde{L}}}$.
\begin{eqnarray}\label{e37}
l&&=\tilde{l}-(\frac{1}{4}(\tilde{b}+1)e\sin 2\tilde{l}\nonumber\\&&+(\frac{1}{64}(\tilde{b}^2+6\tilde{b}+1)e^2 \sin 4 \tilde{l}+O(e^3),
\end{eqnarray}

\begin{eqnarray}\label{e38}
g&&=\tilde{g}+\sqrt{\tilde{b}}\Bigg[\frac{1}{2}e \sin 2\tilde{l}\nonumber\\&&-\frac{1}{16}(\tilde{b}+1)
e^2 \sin 4\tilde{l}\Bigg]+O(e^3)
\end{eqnarray}

\begin{eqnarray}\label{e39}
&&J=\tilde{J}+\frac{1}{16}(2\tilde{b}+1)e^2 \tan \tilde{J}\nonumber\\&&+\tan \tilde{J} (\frac{1}{2} e \cos 2\tilde{l}-\frac{1}{16}\tilde{b}e^2\cos 4\tilde{l})+O(e^3),
\end{eqnarray}

with : 

\begin{eqnarray}\label{e40}
&&\tilde{J}=j+\frac{1}{2}e\tan j+e^2 \tan j (\frac{1}{8}+\frac{3}{16}\tan^2j)+O(e^3)\nonumber\\&&
\tilde{l}=\tilde{n_{l}}\times t \quad \mathrm{with} \quad \tilde{n_{l}}=\frac{G}{D} \cos\tilde{J}[1-\frac{1}{8}(\tilde{b}^2+3)e^2]+O(e^4)\nonumber\\&&
\tilde{g}=\tilde{n_{g}}\times t \quad \mathrm{with}\nonumber\\&&
 \quad \tilde{n_{g}}=\frac{1}{2}(\frac{1}{A}+\frac{1}{B})G+\frac{G}{4D}(\tilde{b}+1)e^2+O(e^4)
\end{eqnarray}

\subsection{Development for a small value of the angle j (Kinoshita, 1972)}
The polar angles $l$ and $J$ leading to the determination of the free rotational motion are given by :
\begin{eqnarray}\label{e41}
l&&=l^{*}-\frac{1}{4}e\sqrt{\frac{1+e}{1-e}}j^2 \frac{\sin 2 \tilde{l}}{1+e \cos 2\tilde{l}}+O(j^4)\nonumber\\&&
g=\tilde{g}+\frac{G}{Dn_{\tilde{l}}}\Big(-\sqrt{1-e^2}(l^{*}-\tilde{l})\nonumber\\&&
+\frac{1}{4}(1+e)j^2\big[\frac{e\sin 2\tilde{l}}{1+e\cos 2\tilde{l}}+\frac{2}{\sqrt{1-e^2}}(l^{*}-\tilde{l})\big]\Big)+O(j^4)\nonumber \\&&
J=j\sqrt{1+\frac{2e}{1-e}\cos^2\tilde{l}}+O(j^3)
\end{eqnarray}
with

\begin{eqnarray}\label{e42}
\tan l^{*}&&=\sqrt{\frac{1-e}{1+e}}\tan \tilde{l}\nonumber\\&&
n_{\tilde{l}}=\frac{G}{D}\sqrt{(1-e^2)}[1-\frac{1}{2(1-e)}j^2]+O(j^4)\nonumber\\&&
n_{\tilde{g}}=\frac{1}{2}(\frac{1}{A}+\frac{1}{B})G+\frac{G}{D}(1-\sqrt{1-e^2})\nonumber\\&&
\times[1+\frac{1}{2}\sqrt{\frac{1+e}{1-e}j^2}]+O(j^4).
\end{eqnarray}


\end{document}